\documentclass[useAMS]{mn2e} 

\usepackage{latexsym,graphicx}

\usepackage{color}

\newcommand\kms{{\rm\,km\,s^{-1}}}
\newcommand\msun{\rm\,M_\odot}

\newcommand\lsun{\rm\,L_\odot}
\newcommand\hii{H\,{\sc ii} \,}

\newcommand\myr{\msun \, {\rm yr}^{-1}}
\newcommand{\MC}{\multicolumn}

\def\apgt{\ {\raise-.5ex\hbox{$\buildrel>\over\sim$}}\ }
\def\aplt{\ {\raise-.5ex\hbox{$\buildrel<\over\sim$}}\ }

\title[HD\,93795 and its circumstellar nebula]{HD\,93795: a late-B supergiant star with a square circumstellar nebula}
\author[V. V.~Gvaramadze et al.] 
       {V. V.~Gvaramadze,$^{1,2}$\thanks{E-mail: vgvaram@mx.iki.rssi.ru} A. Y.~Kniazev,$^{3,4,1}$ 
       N.~Castro$^5$ and I. Y.~Katkov$^{1,6}$ \\
        $^1$Sternberg Astronomical Institute, Lomonosov Moscow State University, Universitetskij Pr. 13, 
        Moscow 119234, Russia\\
        $^2$Space Research Institute, Russian Academy of Sciences, Profsoyuznaya 84/32, 117997 Moscow, Russia \\
        $^3$South African Astronomical Observatory, PO Box 9, 7935 Observatory, Cape Town, South Africa \\
        $^4$Southern African Large Telescope Foundation, PO Box 9, 7935 Observatory, Cape Town, South Africa \\
        $^5$Leibniz-Institut f\"ur Astrophysik Potsdam (AIP), An der Sternwarte 16, 14482, Potsdam, Germany \\
        $^6$New York University Abu Dhabi, Saadiyat Island, PO Box 129188, Abu Dhabi, UAE \\  
        }
\begin{document}

\date{Accepted 2019 December 21. Received 2019 December 18; in original form 2019 December 04}


\maketitle

\label{firstpage}

\begin{abstract}
We report the discovery of a square axisymmetric circumstellar nebula around the emission-line star 
HD\,93795 in archival {\it Spitzer Space Telescope} 24\,$\micron$ data. We classify HD\,93795 as 
an B9\,Ia star using optical spectra obtained with the Southern African Large Telescope (SALT). 
A spectral analysis carried out with the stellar atmosphere code {\sc fastwind} indicates that 
HD\,93795 only recently left the main sequence and is evolving redward for the first time. We discuss 
possible scenarios for the origin of the nebula and suggest that HD\,93795 was originally a binary 
system and that the nebula was formed because of merger of the binary components. We also discuss a 
discrepancy between distance estimates for HD\,93795 based on the {\it Gaia} data and the possible 
membership of this star of the Car\,OB1 association, and conclude that HD\,93795 could be at the same 
distance as Car\,OB1.
\end{abstract}


\section{Introduction}
\label{sec:intro}

\begin{figure*}
\begin{center}
\includegraphics[width=18cm,angle=0]{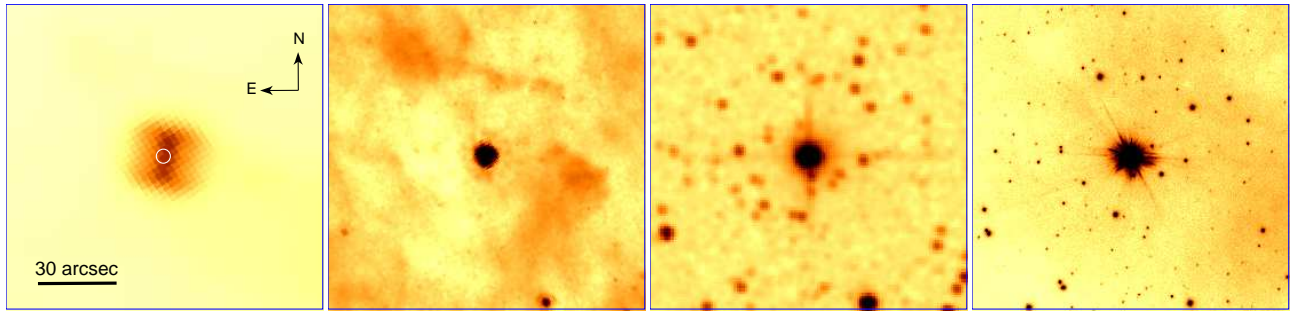}
\end{center}
\caption{From left to right: {\it Spitzer} MIPS 24\,$\micron$ and IRAC 8\,$\micron$, 2MASS 
$K_{\rm s}$ and VPHAS+ H\,$\alpha$ images of HD\,93795 and its circumstellar nebula. A white 
circle in the MIPSE image shows the position of HD\,93795. The orientation and the scale of 
the images are the same. At the distance of 5 kpc, 30 arcsec correspond to $\approx0.72$ pc.
    }
\label{fig:neb}
\end{figure*}

Inspection of archival data from infrared (IR) sky surveys carried out with the {\it Spitzer 
Space Telescope} (Werner et al. 2004) and the {\it Wide-field Infrared Survey Explorer} 
({\it WISE}; Wright et al. 2010) led to the discovery of numerous compact ($\la1$ arcmin 
in diameter) nebulae of various shapes (e.g. Gvaramadze, Kniazev \& Fabrika 2010; Gvaramadze et 
al. 2012). Spectroscopic follow-up of central stars of these nebulae revealed that many of them 
belong to rare types of massive stars, such as the luminous blue variables and Wolf-Rayet stars, 
displaying rich emission spectra (e.g. Kniazev \& Gvaramadze 2015; Gvaramadze \& Kniazev 2017 and 
references therein). A significant part of the nebulae, however, was found to be associated with 
blue supergiant stars whose spectra are either free of emission lines or have mostly the H$\alpha$ 
line in emission. A spectral analysis of some of these stars showed that they only recently left the 
main sequence, meaning that massive stars can produce circumstellar nebulae not only at advanced, 
but also during early stages of their evolution (e.g. Lamers et al. 2001; Smartt et al. 2002; 
Hendry et al. 2008; Gvaramadze et al. 2014, 2015, 2019a; Mahy et al. 2016).

Although the mechanisms responsible for the origin and shaping of circumstellar nebulae are not fully
understood, the axial or bipolar geometry of many of them strongly suggests that binarity or rapid 
rotation of their underlying stars can play a decisive role in their formation.
Detection of new examples of axially-symmetric circumstellar nebulae associated with relatively 
unevolved massive stars could provide important clues about their origin. In this paper, we report the 
discovery of such a nebula around the late-B supergiant star HD\,93795.

In Section\,\ref{sec:neb}, we review the existing data on HD\,93795 and present the nebula. 
Section\,\ref{sec:obs} describes our long-slit and fibre-fed \'echelle spectroscopic observations of 
HD\,93795. The spectral classification of HD\,93795 and the results of modelling of its spectrum with 
the stellar atmosphere code {\sc fastwind} are given in Section\ref{sec:spe}. In 
Section\,\ref{sec:neb-spec}, we report the detection of shocked gas within and beyond the confines 
of the nebula. In Section\,\ref{sec:dis}, we discuss possible mechanisms for the origin of the nebula. 
We summarise in Section\,\ref{sec:sum}.

\section{A square infrared nebula around HD\,93795}
\label{sec:neb}

HD\,93795 is an emission-line star (MacConnell 1981) located in the direction of the giant \hii 
region NGC\,3372 (also known as Carina Nebula), hosting the famous luminous blue variable star 
$\eta$\,Car and several massive star clusters (e.g. Feigelson et al. 2011). HD\,93795 was previously 
classified as A0\,Ia (Hoffleit 1956), B6\,I (Stephenson \& Sanduleak 1971), B9\,Iab (Humphreys 1973), 
and A2/3p: shell: (Houk \& Cowley 1975). 

The line of sight to HD\,93795 is tangential to the Sagittarius-Carina spiral arm, meaning that 
this star is located somewhere between $\approx2$ and 7 kpc from the Sun. According to Humphreys 
(1978), HD\,93795 is a member of the Car\,OB1 association, which is located at the distance of 
$2.3\pm0.2$ kpc (Lim et al. 2019). The {\it Gaia} second data release (DR2; Gaia Collaboration 
et al. 2018) parallax of HD\,93795 of $0.1975\pm0.0356$\,mas places this star at a distance of 
$d=5.06^{+1.11} _{-0.77}$ kpc. A somewhat shorter distance to HD\,93795 of $d=4.32^{+0.80} _{-0.59}$ 
kpc was derived from the DR2 data by Bailer-Jones et al. (2018) by using a Bayesian prior knowledge 
approach. Both these distances are much larger than the distance to Car\,OB1, suggesting that 
HD\,93795 might be a background star projected by chance on Car\,OB1. In what follows, we adopt a 
distance of $d=5\pm1$ kpc for HD\,93795. Still, due to the zero-point problem in the {\it Gaia} DR2 
parallaxes (Lindegren et al. 2018; Stassun \& Torres 2018; Zinn et al. 2019) and because the parallax 
measurements could be affected by possible binarity of HD\,93795 (suggested by radial velocity 
variability of the star; see below and Section\,\ref{sec:mod}), one cannot exclude the possibility 
that this star is located at the same distance as Car\,OB1 (cf. Sections\,\ref{sec:lum} and 
\ref{sec:neb-spec}). We will consider this possibility as well.

In literature we found three measurements of the heliocentric radial velocity, $v_{\rm r,hel}$, of 
HD\,93795. Two of them, $-20.7\pm2.2 \, \kms$ and $-4.1\pm1.5 \, \kms$, are given in Humphreys (1973), 
and the third one, $-12.4 \, \kms$, in Humphreys (1978). We compare them with our own measurements 
and discuss in Section\,\ref{sec:mod}.

In Table\,\ref{tab:det}, we provide a short summary of the properties of HD\,93795. The spectral type 
is based on our observations. The coordinates and the $B$ and $V$ 
magnitudes are from the Tycho Input Catalogue (Egret et al. 1992). The $J$ and $K_{\rm s}$ magnitudes 
are from the Two-Micron All Sky Survey (2MASS; Skrutskie et al. 2006) All-Sky Catalog of Point Sources 
(Cutri et al. 2003). 

\begin{table}
  \centering{\caption{Properties of HD\,93795.}
  \label{tab:det}
 \begin{tabular}{lc}
    \hline
  Spectral type & B9\,Ia \\
  RA(J2000) & $10^{\rm h} 48^{\rm m} 23\fs10$  \\ 	
  Dec.(J2000) & $-59\degr 32\arcmin 26\farcs9$ \\ 
  $l$ & $287\fdg9038$ \\
  $b$ & $-0\fdg3083$  \\
  $B$ (mag) & $9.30\pm0.01$ \\
  $V$ (mag) & $8.50\pm0.01$ \\
  $J$ (mag) & $6.43\pm0.02$ \\  
  $K_{\rm s}$ (mag) & $5.90\pm0.04$ \\
  \hline
 \end{tabular}
}
\end{table}

The nebula around HD\,93795 was discovered in the search for bow-shock-producing stars running away
from massive star clusters, including those in the Carina Nebula (for motivation of this search see 
Gvaramadze \& Bomans 2008). We found it in the MIPS (Multiband Imaging Photometer for {\it Spitzer}; 
Rieke et al. 2004) 24\,$\micron$ image obtained in the framework of the {\it Spitzer} survey of the 
Carina Nebula (program ID 30848; N.\,Smith, PI). The nebula is not visible at shorter wavelengths, 
nor in the IRAC (Infrared Array Camera; Fazio et al. 2004) 8, 5.8, 4.5 and 3.6\,$\micron$ images 
from the {\it Spitzer} Vela-Carina survey (program ID 40791; S.R.\,Majewski, PI), nor in the 
H\,$\alpha$ images from the SuperCOSMOS H-alpha Survey (Parker et al. 2005) and the VST Photometric 
H\,$\alpha$ Survey of the Southern Galactic Plane and Bulge (VPHAS+; Drew et al. 2014). 

In the 24\,$\micron$ image the nebula appears as a square of $\approx30$ arcsec side with a bright
ridge in the middle and two symmetric flanks perpendicular to the ridge (see Fig.\,\ref{fig:neb}).
The linear size of the nebula scales with the distance as $\approx0.72(d/5 \, {\rm kpc})$\,pc.
The geometry of the nebula suggests that we deal with a disc-like structure viewed almost edge-on 
(the ridge) and a bipolar outflow perpendicular to the disc plane. In this connection, we note that 
Heiles (2000) measured polarization towards HD\,93795 of $2.21\pm0.40$ per cent with a position 
angle of PA=$95.0\degr\pm5.2\degr$ (measured from north to east). If a significant fraction of the 
polarization is intrinsic to the star itself, then this would imply the presence of a flattened 
circumstellar structure around the star with the major axis roughly parallel to the ridge in the IR 
nebula (whose PA is $\approx168\degr$).

\begin{figure*}
\begin{center}
	\includegraphics[width=12cm]{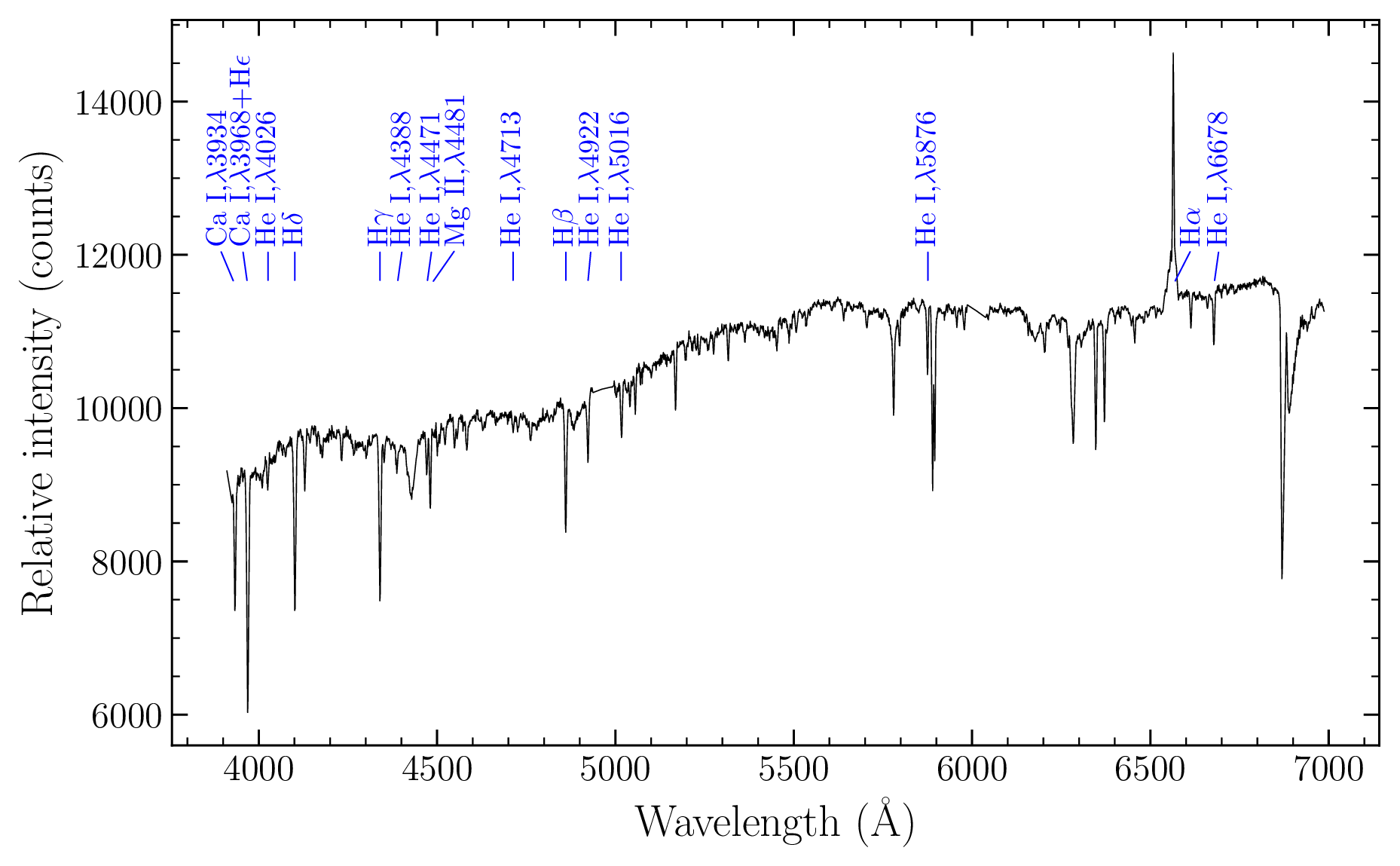}
	\caption{The observed RSS spectrum of HD\,93795 obtained on 2015 July 3, with the main
	lines indicated. The gaps in the spectrum at $\approx4900$ and 6000\,\AA \, are due to gaps 
	between CCDs.
	}
	\label{fig:rss}
	\end{center}
\end{figure*}

\section{Spectroscopic observations}
\label{sec:obs}

We observed HD\,93795 for the first time on 2013 April 26 as part of our ongoing program of 
spectroscopic follow-up of central stars of IR nebulae detected with {\it Spitzer} and {\it WISE} 
(e.g. Gvaramadze et al. 2010, 2012). The observation was carried out with the Robert Stobie 
Spectrograph (RSS; Burgh et al. 2003; Kobulnicky et al. 2003) at the Southern African Large 
Telescope (SALT; Buckley, Swart \& Meiring 2006; O'Donoghue et al. 2006). The PG900 grating was 
used along with a 8 arcmin$\times$1.25~arcsec slit, covering the spectral range of 4200--7300 \AA\ 
with a final reciprocal dispersion of 0.97\,\AA\ pixel$^{-1}$. The spectral resolution FWHM was 
4.47$\pm$0.25~\AA. Three science spectra were taken with exposure times of 1, 10 and 60s to avoid 
possible saturation of the H\,$\alpha$ emission line. An Xe lamp arc spectrum was taken immediately 
after the science frame. The slit was oriented at PA=170$\degr$. The seeing during the observations 
was 1.5~arcsec.

HD\,93795 was observed for the second time with RSS on 2015 July 3. Again the PG900 grating was 
used with the 8 arcmin$\times$1.25~arcsec slit, but this time with the spectral setup covering 
the spectral range of 3900--7000 \AA\ with a final reciprocal dispersion of 0.97\,\AA\ pixel$^{-1}$.
The spectral resolution FWHM was 4.45$\pm$0.23~\AA. Two spectra were taken with exposures of
15 and 180\,s (in both cases the slit was oriented at PA=80$\degr$). The seeing during these
observations was about 2.0~arcsec. The short exposure was used to search for possible major changes 
in the spectrum of the star, while the longer exposure was intended to search for nebular emission 
around the star.

A spectrum of an Ar comparison arc was obtained immediately after the observations to calibrate 
the wavelength scale. For the relative flux calibration the spectrophotometric standard stars EGGR\,21 
(Baldwin \& Stone 1984) and LTT\,9239 (Hamuy et al.\, 1992, 1994) were observed with the same spectral 
set-ups during nearest twilights as a part of the SALT calibration plan. 

The RSS spectra were first reduced with the SALT science pipeline (Crawford et al. 2010) and then 
further reduced in the way described in Kniazev et al. (2008).
The spectrum obtained in 2013 was presented in Kniazev \& Gvaramadze (2015), where we tentatively 
classified HD\,93795 as an A star (no luminosity class was derived at that time). The only emission 
line detected in this spectrum is H\,$\alpha$. The short-exposure spectrum of HD\,93795 from 2015 
is presented in Fig.\,\ref{fig:rss}. The comparison of the RSS spectra showed that the intensity of 
the H\,$\alpha$ line has increased in 2015 by about 10 per cent, while its overall shape is slightly 
changed (cf. Section\,\ref{sec:mod}). The analysis of the long-exposure spectrum led to the 
detection of shocked gas within and beyond the confines of the IR nebula (see 
Section\,\ref{sec:neb-spec}).

For spectral modelling and to search for possible radial velocity variability, we obtained two 
spectra of HD\,93795 with the High Resolution Spectrograph (HRS; Barnes et al. 2008; Bramall et al. 
2010, 2012; Crause et al. 2014), which is also mounted on SALT. The spectra were taken in the 
medium resolution (MR) mode on 2018 December 21 and 2019 February 5 with the exposure time of each 
observation of 210\,s. The seeing during these observations was 1.8 and 1.7 arcsec, respectively. 

The HRS is a dual beam, fibre-fed \'echelle spectrograph. In the MR mode it has 2.23 arcsec 
diameter for both the object and sky fibres providing a spectrum in the blue and red arms over 
the spectral range of $\approx$3700--8900~\AA\ with resolving power of $R\approx40\,000$ 
(Kniazev et al. 2019). Both the blue and red arm CCDs were read out by a single 
amplifier with a 1$\times$1 binning. A spectrum of a ThAr lamp and three spectral flats were 
obtained in the MR mode during a weekly set of HRS calibrations. For the relative flux calibration
spectrophotometric standard stars from the list in Kniazev (2017) were observed once per week as a 
part of the SALT HRS calibration plan. During our observations the star ksi02 Cet (Turnshek et al. 
1990) was used as a standard.

The spectra were first reduced with the SALT science pipeline (Crawford et al. 2010) and then
further reduced as described in Kniazev, Gvaramadze \& Berdinikov (2016).

We also retrieved and co-added processed spectra of HD\,93795 from the European Southern Observatory 
(ESO) data archive. These spectra were obtained on 2014 April 26 under programme ID 092.A-9020(A) with 
the high-resolution ($R=48\,000$) \'echelle spectrograph FEROS (Fibre-fed Extended Range Optical 
Spectrograph; Kaufer et al. 1999) at the 2.2-m Max Planck Gesellschaft telescope (La Silla). Eight 
spectra covering the spectral range of $\approx3500-9200$~\AA \, were taken with a 
total exposure time of 2400\,s. The seeing in these observations was $\approx1.4-1.6$ arcsec. The 
obtained spectra were reduced with the ESO pipeline.

\section{Spectral analysis}
\label{sec:spe}

\subsection{Spectral classification}
\label{sec:cla}

The spectrum of HD\,93795 is dominated by the H\,{\sc i} and He\,{\sc i} lines (see 
Fig.\,\ref{fig:rss}). The H\,$\alpha$ line is in emission and shows a complex profile (see below). 
In the HRS spectra there are several weak emission lines of Fe\,{\sc ii} at 7496, 7513, 7732, 7755, 
7789 and 7848 \AA. The Mg\,{\sc ii} $\lambda$4481/He\,{\sc i} $\lambda$4471 and Fe\,{\sc ii} 
$\lambda$4233/Si\,{\sc ii} $\lambda\lambda$4128, 4132 intensity ratios of $>1$ and $<1$, respectively,
imply that HD\,93795 is of B9 spectral type (see Evans et al. 2004), which agrees with the spectral
classification by Humphreys (1973). 

To derive the luminosity class, we use the Keenan \& Hynek (1950) calibration between equivalent 
width (EW) of the O\,{\sc i} $\lambda\lambda$7772--5 \AA \, triplet and the luminosity class. With 
EW(O\,{\sc i})=$1.80\pm0.01$ \AA \, (measured in the HRS spectra), one finds the luminosity class Ia, 
meaning that HD\,93795 is more luminous then it is suggested by the luminosity class Iab given in 
Humphreys (1973, 1978).

\begin{figure*}
\begin{center}
	\includegraphics[width=18cm]{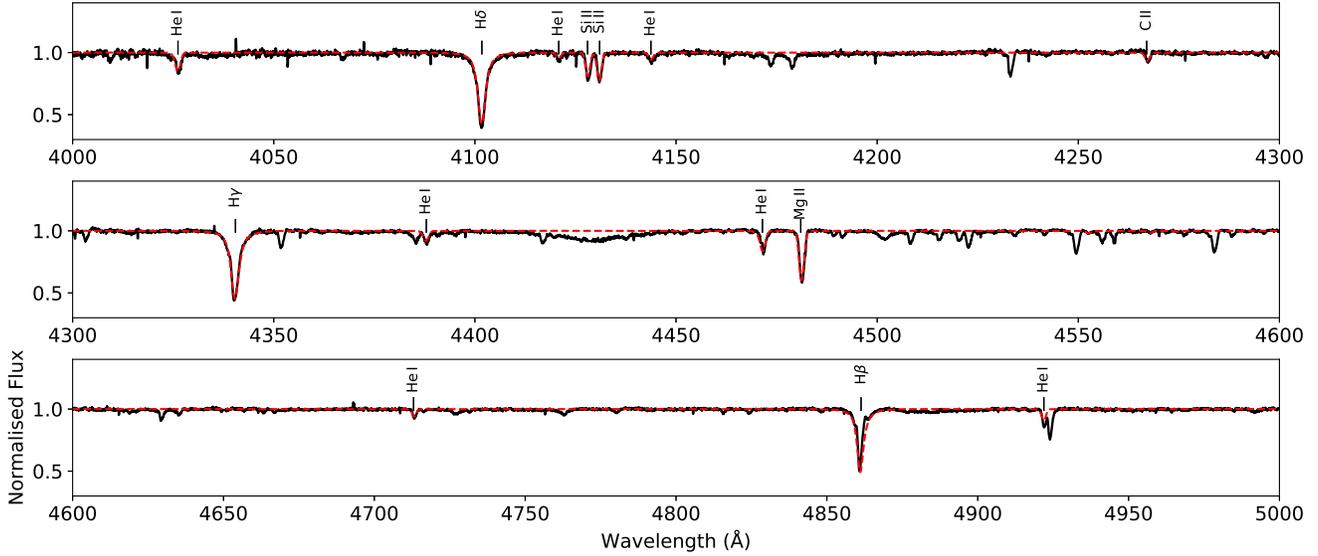}
	\caption{Parts of the normalized and rebinned spectrum of HD\,93795 (black line), compared 
	with the best-fitting {\sc fastwind} model (red dashed line) with the parameters as given in 
	Table\,\ref{tab:par}. The lines fitted by the model are indicated.
	}
	\label{fig:mod}
	\end{center}
\end{figure*}

\subsection{Spectral modelling and stellar parameters}
\label{sec:mod}

The three spectra of HD\,93795 were corrected for heliocentric velocity and radial velocity variations 
(see below), and co-added to increase the signal-to-noise ratio. The resulting spectrum 
was analysed using the stellar atmosphere code {\sc fastwind} (Santolaya-Rey, Puls \& Herrero 1997; 
Puls et al. 2005; Rivero Gonz\'alez et al. 2012). {\sc fastwind} calculates the atmosphere and 
line formation in spherical symmetry, with an explicit treatment of the stellar wind and
taking into account non-local thermodynamic equilibrium (non-LTE).

The quantitative analysis was carried out in two steps. First, several key lines of H\,{\sc i}, 
He\,{\sc i} and Si\,{\sc ii} were compared to a synthetic {\sc fastwind} grid tailored for the 
analysis of B-type supergiants. The quantitative analysis follows the $\chi^2$ approach described 
in Castro et al. (2012). A full description of the stellar parameters covered in the grid and the 
list of lines used in the analysis can also be found in Castro et al. (2012). 

Second, and based on the stellar parameters obtained in the first step (see Table\,\ref{tab:par}),
a new {\sc fastwind } sub-grid was build for different chemical abundances of Mg, C, Si, O and N. 
The grid ranged $\pm2\,$dex around the cosmic abundance standard (CAS) in the solar neighborhood 
(Nieva \& Przybilla 2012) in steps of 0.1\,dex. The grids was also built for different 
microturbulence velocities, $v_{\rm mic}$, exploring values from 5 to $25 \,\kms$ in steps of $1 
\, \kms$. Different chemical abundances and microturbulence velocities were simultaneously examined 
until finding the best match with the observations. The best-fitting model is displayed in 
Fig.~\ref{fig:mod}, while the stellar parameters derived from this model are compiled in 
Table~\ref{tab:par}. Note that at the effective temperature of HD\,93795, we could only constrain 
the chemical abundances of those elements with strong absorption lines whose transitions were 
included in our models, i.e. Mg, C and Si. The main lines used for the spectral modelling are marked 
in Fig.\,\ref{fig:mod}.

\begin{table}
\caption{Stellar parameters and non-LTE metal abundances (by number) of HD\,93795.
The cosmic abundance standard in the solar neighbourhood (Nieva \& Przybilla 2012) is 
given in brackets for reference.}
\label{tab:par}
\begin{center}
\begin{tabular}{lc}
\hline
$T_\mathrm{eff}$ (kK)         & $11\pm1$  \\
$\log g$ (cgs)                & $1.6\pm0.1$      \\
$v_{\rm mic} \, (\kms)$       & $13\pm2$          \\
$v_{\rm mac} \, (\kms)$       & $50\pm10$         \\
$v\sin i \, (\kms)$           & $55\pm8$          \\	
$\log$\,(He/H)\,$+$\,12       & $11.0\pm0.20$ ($10.99\pm0.01$) \\
$\log$\,(C/H)\,$+$\,12        & $8.0\pm0.15$ ($8.33\pm0.04$) \\
$\log$\,(Mg/H)\,$+$\,12       & $7.6\pm0.10$ ($7.56\pm0.05$) \\
$\log$\,(Si/H)\,$+$\,12       & $7.5\pm0.10$ ($7.50\pm0.05$) \\
\hline
\end{tabular}
\end{center}
\end{table}

\begin{table}
\caption{Changes in the heliocentric radial velocity of HD\,93795.}
\label{tab:rad}
\begin{center}
\begin{tabular}{lc} \hline
Date & $v_{\rm r,hel} (\kms)$ \\
\hline
2014 March 9 & $-9.3\pm0.1$ \\  
2018 December 21 & $-5.5\pm0.3$ \\
2019 February 5 & $-16.5\pm0.2$ \\ 
\hline
\end{tabular}
\end{center}
\end{table}

\begin{figure*}
\begin{center}
\includegraphics[width=6cm,angle=270,clip=]{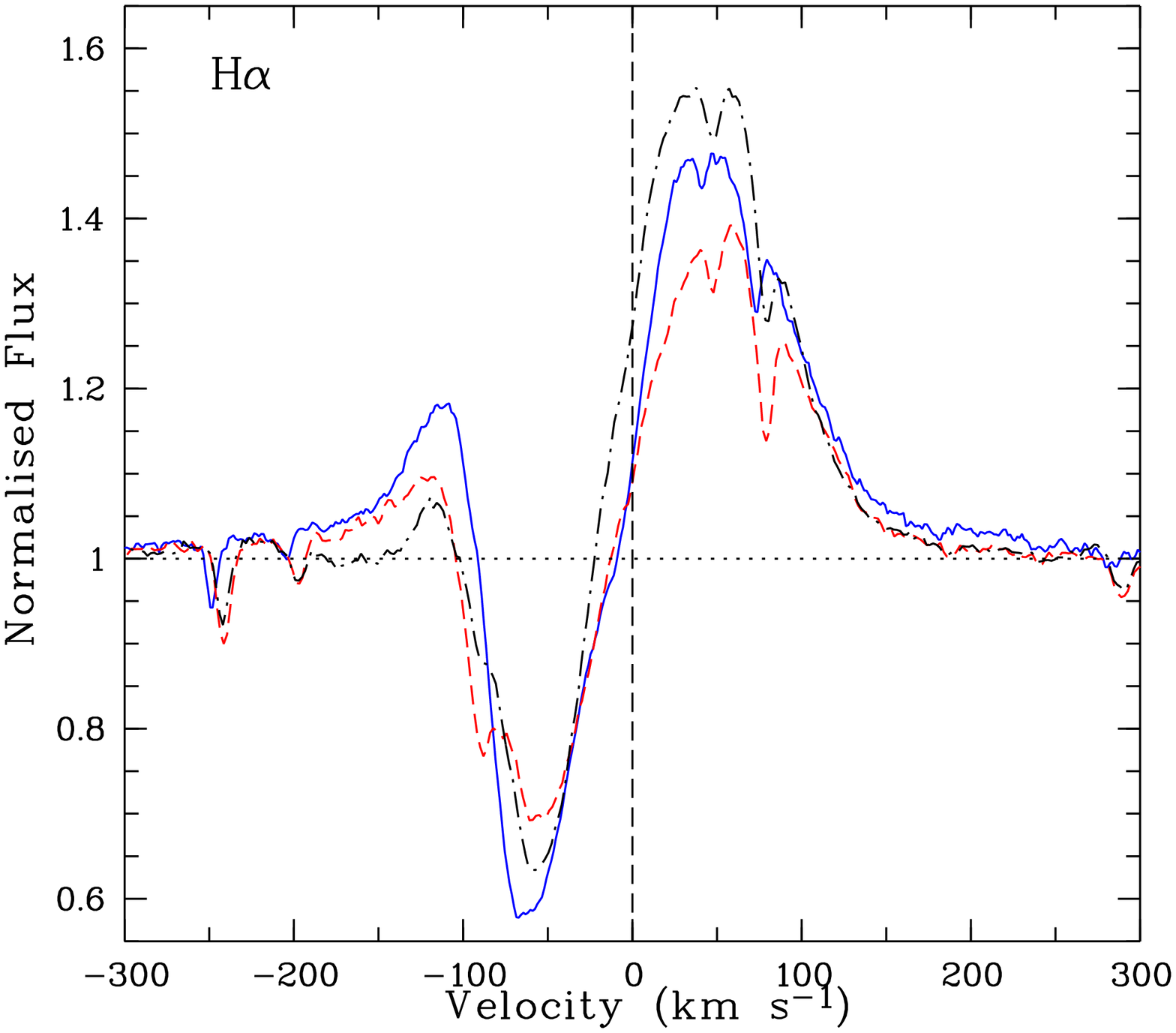} 
\includegraphics[width=6cm,angle=270,clip=]{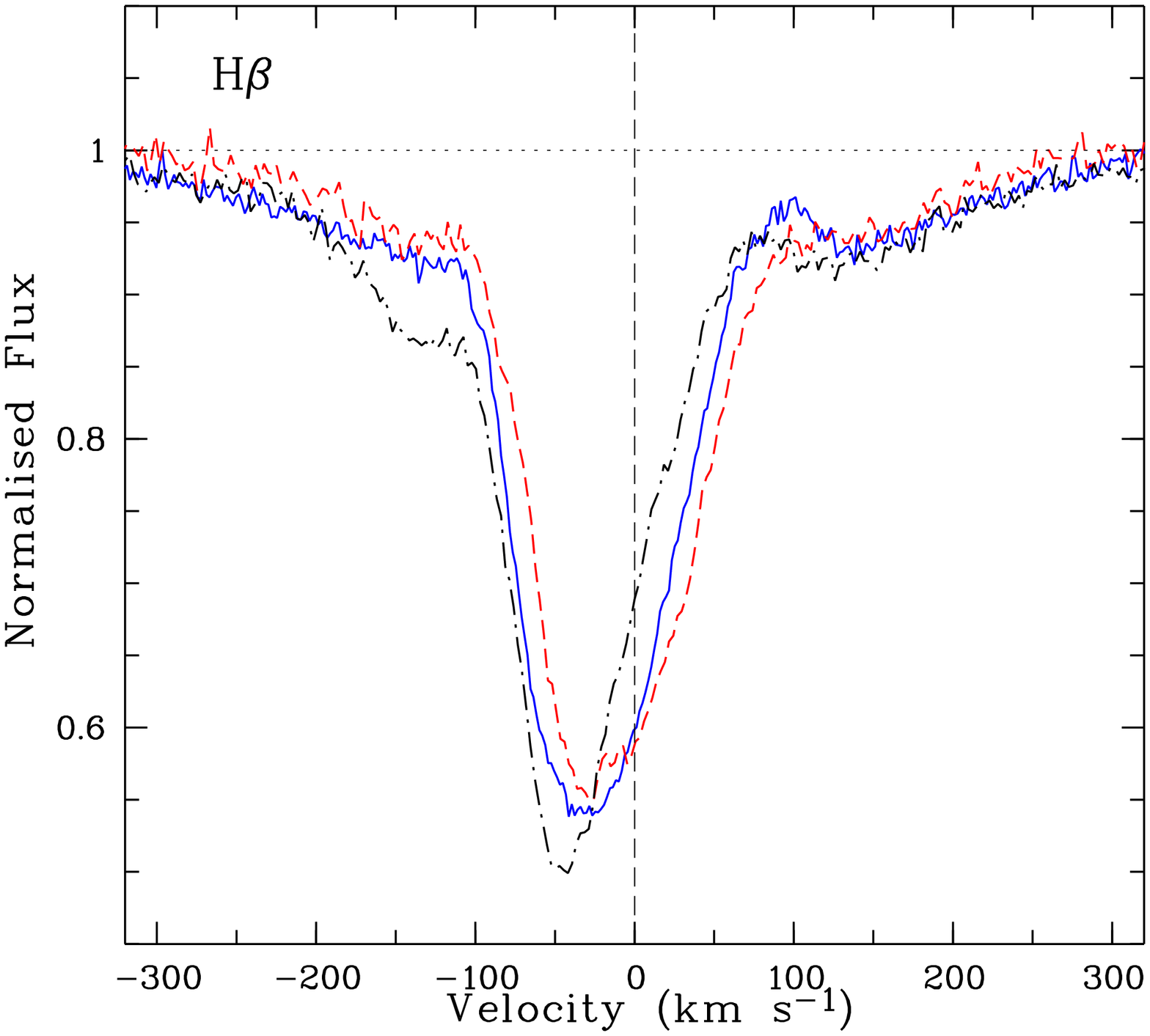} 
\includegraphics[width=6cm,angle=270,clip=]{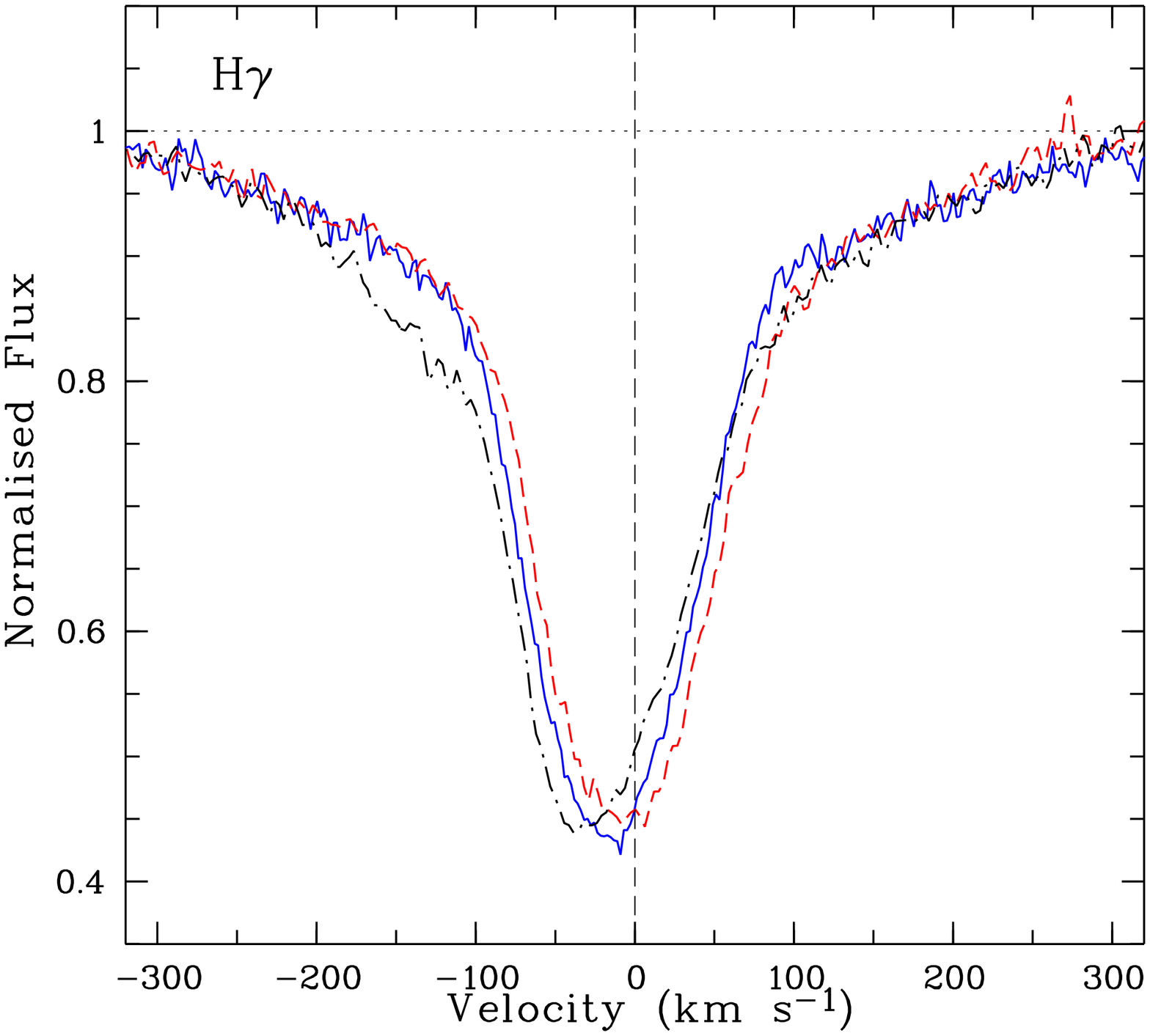} 
\includegraphics[width=6cm,angle=270,clip=]{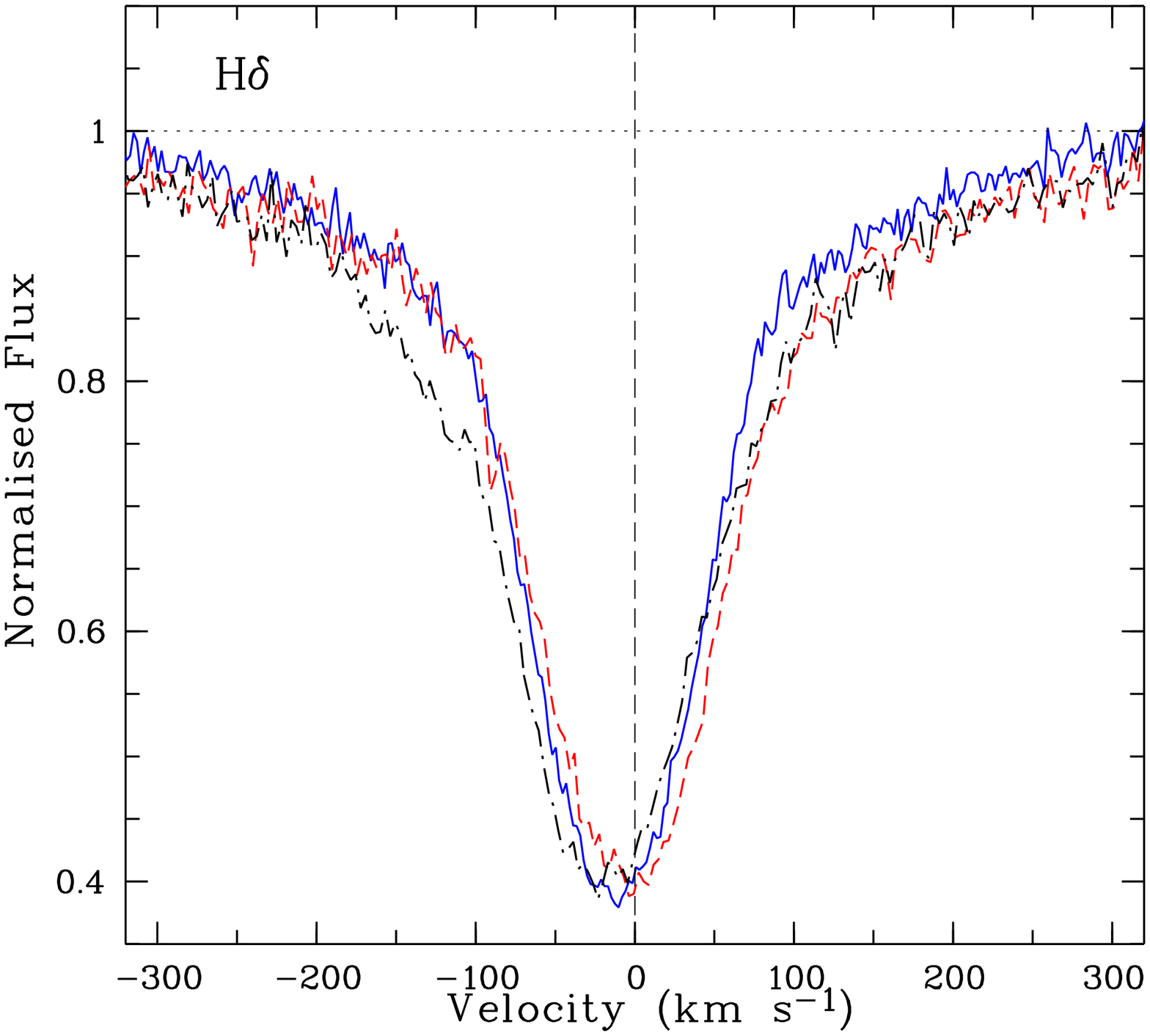} 
	\caption{Changes in the Balmer line profiles in the spectra obtained with FEROS on 2014 March 9 
	(solid blue line) and HRS on 2018 December 21 (dashed red line) and 2019 February 5 (dot-dashed 
	black line). Velocities are relative to the zero heliocentric radial velocity (vertical dashed 
	line), corresponding to the rest wavelengths of the lines. Narrow dips in the H\,$\alpha$ line 
	profiles are due to telluric H$_2$O absorption lines. 
		}
	\label{fig:ha}
\end{center}
\end{figure*}
%
\begin{figure}
\begin{center}
	\includegraphics[width=8cm,clip=]{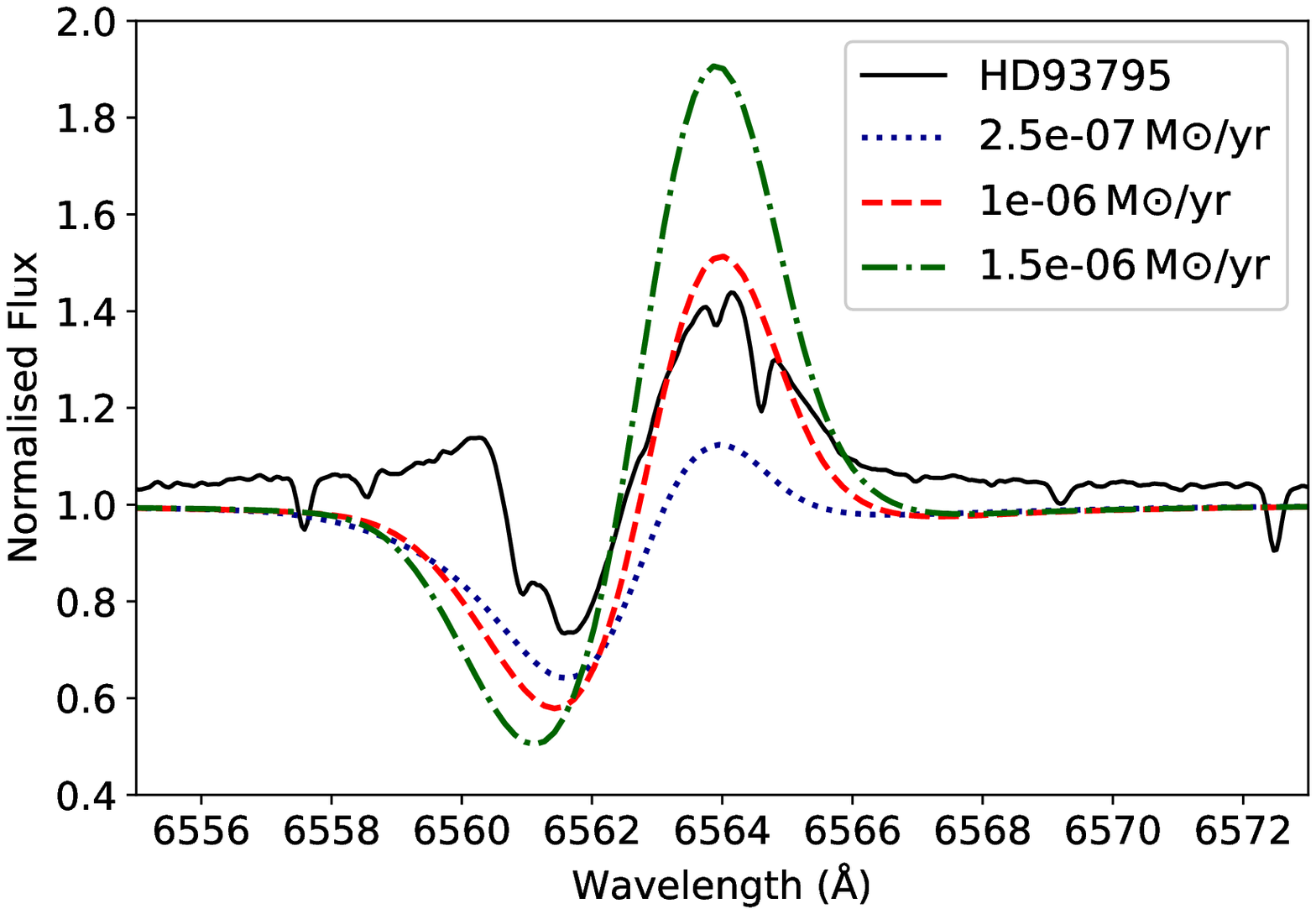}
	\caption{Comparison of the observed profile of the H$\alpha$ line (from the 2019 spectrum)
	with synthetic profiles predicted by {\sc fastwind} models with different values of the 
	mass-loss rate as specified in the legend.
	}
	\label{fig:ha2}
\end{center}
\end{figure}
 
We used the {\sc iacob-broad} code Sim\'on-D\'iaz \& Herrero (2014) to measure the projected rotational 
($v\sin i$) and macroturbulence ($v_{\rm mac}$) velocities. This code characterizes the additional 
broadening in the line profiles combining Fourier transform and goodness-of-fit methodologies. The 
obtained velocities (see Table~\ref{tab:par}) are qualitatively in agreement with velocities 
measured by Sim\'on-D\'iaz et al. (2017) in stars at the similar evolutionary stages (see their 
fig.\,5). 

Using the available three \'echelle spectra, we also measured the heliocentric radial velocity, 
$v_{\rm r,hel}$, of HD\,93795 in the way described in Gvaramadze et al. (2019b). The obtained values 
of $v_{\rm r,hel}$ (see Table\,\ref{tab:rad}) are similar to those measured by Humphreys (1973, 1978), 
who found radial velocity variations in the range between $\approx-4$ and $-21 \, \kms$ (see 
Section\,\ref{sec:neb}). The changes in $v_{\rm r,hel}$ suggest that HD\,93795 might be a binary 
system, but they could also be caused by pulsations of the stellar photosphere (e.g. 
Jeffery \& Saio 2016). Time-series photometry and spectroscopy of HD\,93795 are necessary 
to distinguish between these two possibilities.

Comparison of the three spectra also revealed changes in the Balmer line profiles, which are most 
prominent in the H\,$\alpha$ line (see Fig.\,\ref{fig:ha}). This line shows a distinctive P\,Cyg 
profile (suggesting that the star possesses a strong wind) and also has an emission component in 
the blue wing (note that narrow dips in the line profile are telluric in origin). The H\,$\alpha$ 
line variability was detected in almost all late-B and early-A supergiants (e.g., Kaufer et al.
1996; Markova \& Valchev 2000; Verdugo et al. 2000), which in most cases could be characterised as 
a variable double-peaked emission pattern symmetric about the systemic velocity. By analogy with 
the V/R variability in the H\,$\alpha$ line profiles in Be stars, this variability was interpreted 
(e.g., Kaufer et al. 1996) in terms of an axially-symmetric stellar wind and/or a disc-like 
structure around the star. 

In Fig.\,\ref{fig:ha2}, we compare the observed H\,$\alpha$ line profile from the 2019 spectrum with 
the {\sc fastwind} synthetic profiles obtained for three different mass-loss rates 
$\dot{M}$ (indicated in the legend of the figure). To produce the synthetic profiles, we adopted 
a terminal wind velocity of $v_\infty =220 \, \kms$, based on the effective temperature of HD\,93795
and the empirical calibration in Kudritzki \& Puls (2000), and a wind velocity law exponent
$\beta=3.5$ (see equation (3) in Castro et al. 2012). No clumping was included in the models.

Fig.\,\ref{fig:ha2} shows that although the intensity of the observed line matches a mass-loss rate 
of $\approx10^{-6} \, \myr$, the synthetic profiles did not reproduce the wings of the line.
This discrepancy could be due to shortcomings of our models or/and due to 
non-spherical geometry of the stellar wind (e.g. Kaufer et al. 1996; Kudritzki et al. 1999).

\subsection{Reddening, luminosity and evolutionary status of HD\,93795}
\label{sec:lum}

To determine the $V$-band reddening towards HD\,93795, one can use the photometric data from 
Table\,\ref{tab:det} and the intrinsic colours $(B-V)_0$ and $(J-K)_0$ of B9 supergiants of, 
respectively, $-0.06$\,mag (Firnstein \& Przybilla 2012) and $-0.05$\,mag (Wenger 2014):
\begin{equation}
A_V=R_V[(B-V)-(B-V)_0] 
\label{eq:vis}
\end{equation}
and
\begin{equation}
A_V=5.89[(J-K)-(J-K)_0] \, ,
\label{eq:ir}
\end{equation}
where $R_V=A_V/E(B-V)$ is total-to-selective absorption ratio, $K = K_{\rm s}+0.04$ mag (Carpenter 
2001). To derive equation\,(\ref{eq:ir}), we used the extinction law of Rieke \& Lebofsky (1985). 
From equations\,(\ref{eq:vis}) and (\ref{eq:ir}), one finds $A_V=0.86 R_V$ mag and $A_V=3.18$ mag, 
respectively, which would agree with each other if $R_V\approx3.7$. This suggests that the reddening 
towards HD\,93795 is anomalous, which is in good agreement with the results by Herbst (1976) and 
Forte (1978), showing that the extinction law towards the Carina Nebula is characterized by 
significantly larger values of $R_V$ ($\approx4-5$) then the ``normal" value of 3.1.

Adopting $A_V=3.18$ mag and using the bolometric correction for B9 supergiants of BC=$-0.31$ mag 
(Firnstein \& Przybilla 2012), one finds the absolute $V$-band magnitude and luminosity of HD\,93795 
of
\begin{eqnarray}
M_V=V-5\log(d/{\rm pc}) +5-A_V \approx -8.17^{+0.48} _{-0.40} \, {\rm mag} 
\nonumber
\end{eqnarray}
and
\begin{eqnarray}
\log (L/\lsun)=0.4(4.74-M_V -{\rm BC})\approx5.3\pm0.2 \, , 
\nonumber
\end{eqnarray}
respectively. For the error calculation, only uncertainties in the adopted distance of $5\pm1$ kpc 
were considered. The obtained luminosity implies that the zero-age main-sequence mass of HD\,93795 
was $\approx25\pm5 \, \msun$ (e.g. Ekstr\"om et al. 2012). In the case if the star is located at 
$d=2.3$\,kpc, its luminosity and initial mass would be equal to $\log (L/\lsun)\approx4.6$ and 
$\approx15 \, \msun$, respectively.

The luminosity of HD\,93795 could also be estimated using the flux-weighted gravity-luminosity 
relationship (Kudritzki, Bresolin \& Przybilla 2003):
\begin{equation}
M_{\rm bol}=(3.41\pm0.16)(\log g_{\rm F} -1.5)-(8.02\pm0.04) \, , 
\label{eq:bol}
\end{equation}
where $M_{\rm bol}=M_V+{\rm BC}$ is the bolometric magnitude and $\log g_{\rm F}=\log g 
-4\log(T_{\rm eff} \times10^{-4})$. For $\log g=1.6\pm0.1$ and $T_{\rm eff}=11\pm1$\,kK, one finds 
from equation\,(\ref{eq:bol}) that $M_{\rm bol}=-8.3\pm1.0$\,mag, which corresponds to the stellar 
luminosity of $\log (L/\lsun)=5.2\pm0.4$. Taken at face value, this estimate agrees well with that 
based on the {\it Gaia} distance, but still did not exclude the possibility that HD\,93795 is 
located at a shorter distance.

Our spectral modelling showed that the helium abundance in HD\,93795 is close to the CAS value, 
while the carbon abundance is only slightly depleted (see Table\,\ref{tab:par}). This indicates 
that HD\,93795 only recently left the main sequence and is evolving redward (to cooler temperatures) 
for the first time. Correspondingly, the luminosity and the temperature of HD\,93795 would imply the 
age of this star of $\ga10$\,Myr (Ekstr\"om et al. 2012). This age is much older than the age of the 
central massive star clusters in the Carina Nebula of 1--3 Myr (Hur et al. 2012), but comparable to 
the kinematic age of the Car\,OB1 association of $\sim 10$\,Myr (Melnik \& Dambis 2017).

\begin{figure}
\begin{center}
	\includegraphics[width=8cm]{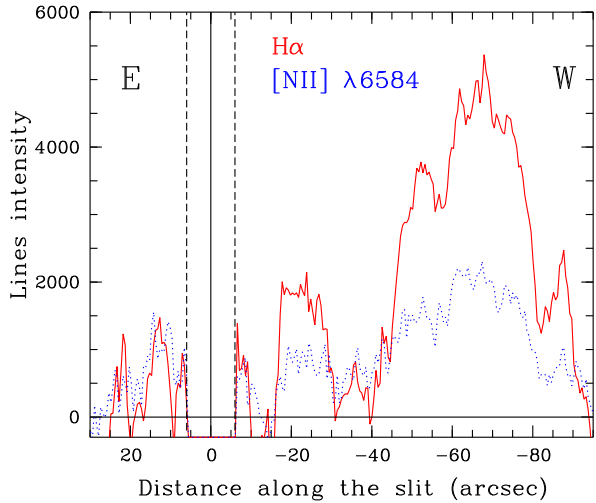}
	\caption{Variations of the H\,$\alpha$ and [N\,{\sc ii}] $\lambda$6584 emission line intensities 
	along the slit with PA=80$\degr$, shown, respectively, with a solid (red) and dashed (blue) lines.
	The solid vertical line corresponds to the position of HD\,93795, while the dashed vertical lines 
	at $\pm6$ arcsec from the solid line mark the area where no nebular emission was detected because 
	of the effect of HD\,93795. E--W direction of the slit is shown. 
	}
	\label{fig:pro}
	\end{center}
\end{figure}

\begin{figure*}
\begin{center}
	\includegraphics[width=8.5cm,clip=]{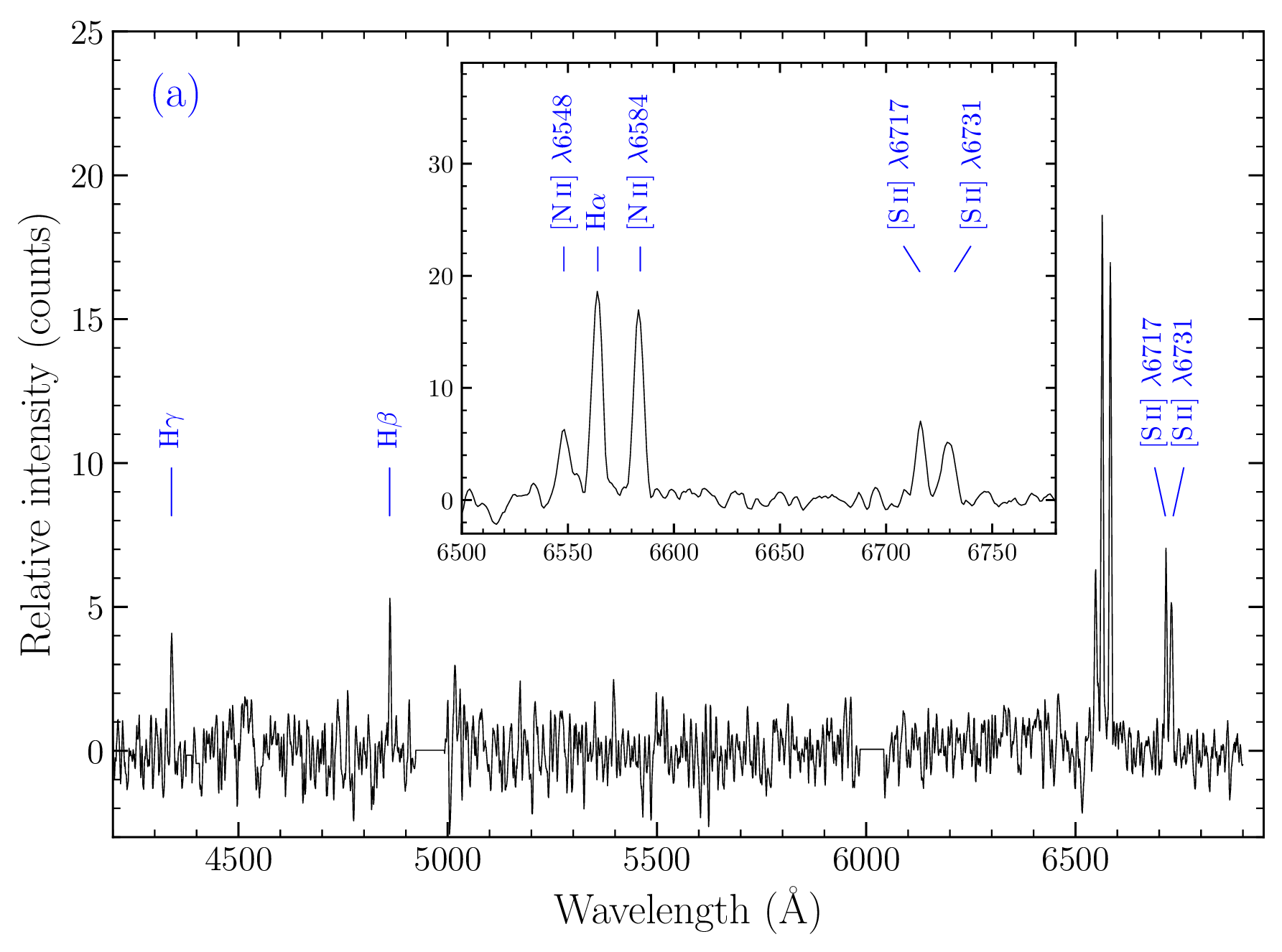}
	\includegraphics[width=8.5cm,clip=]{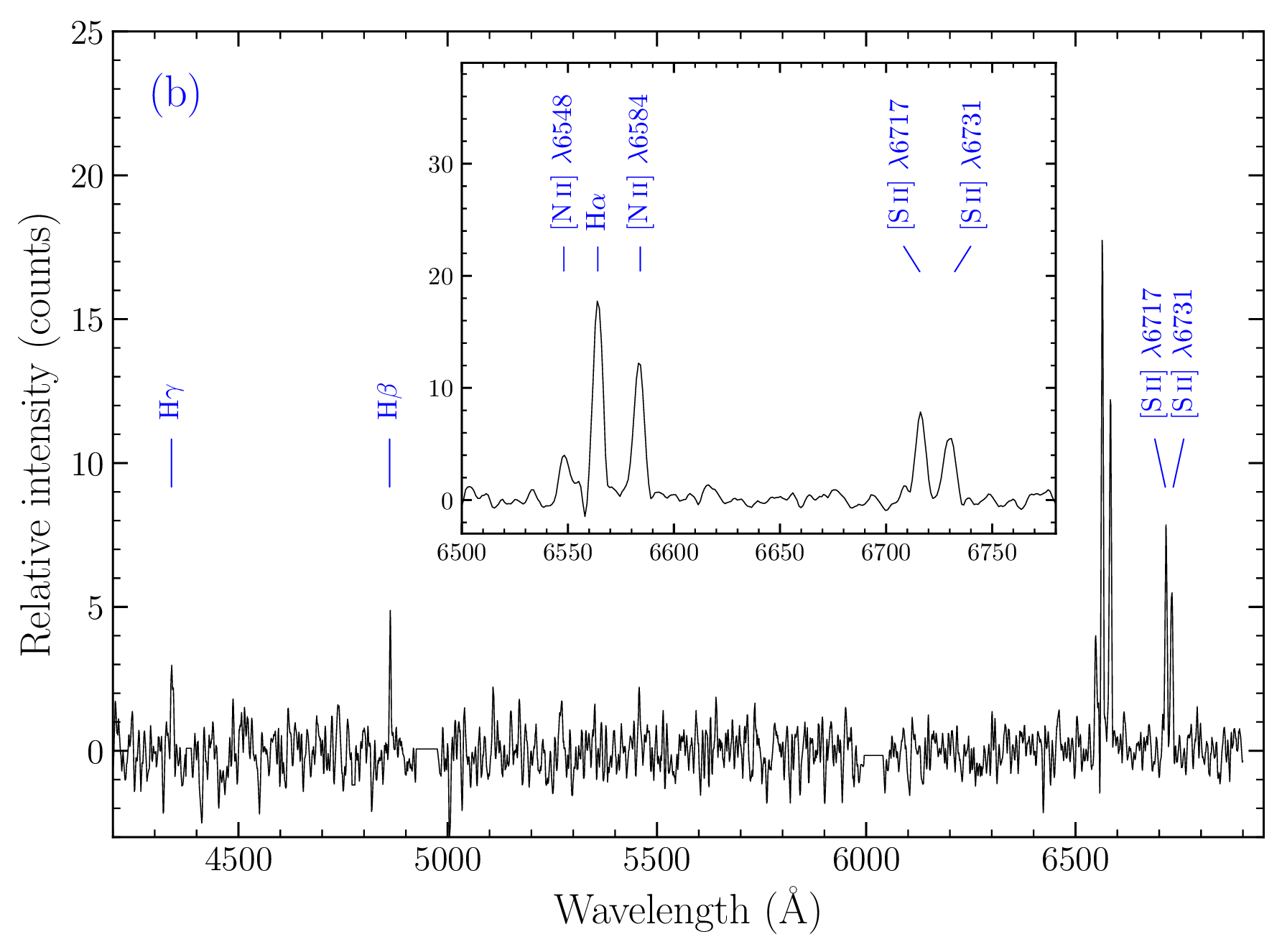}
	\includegraphics[width=8.5cm,clip=]{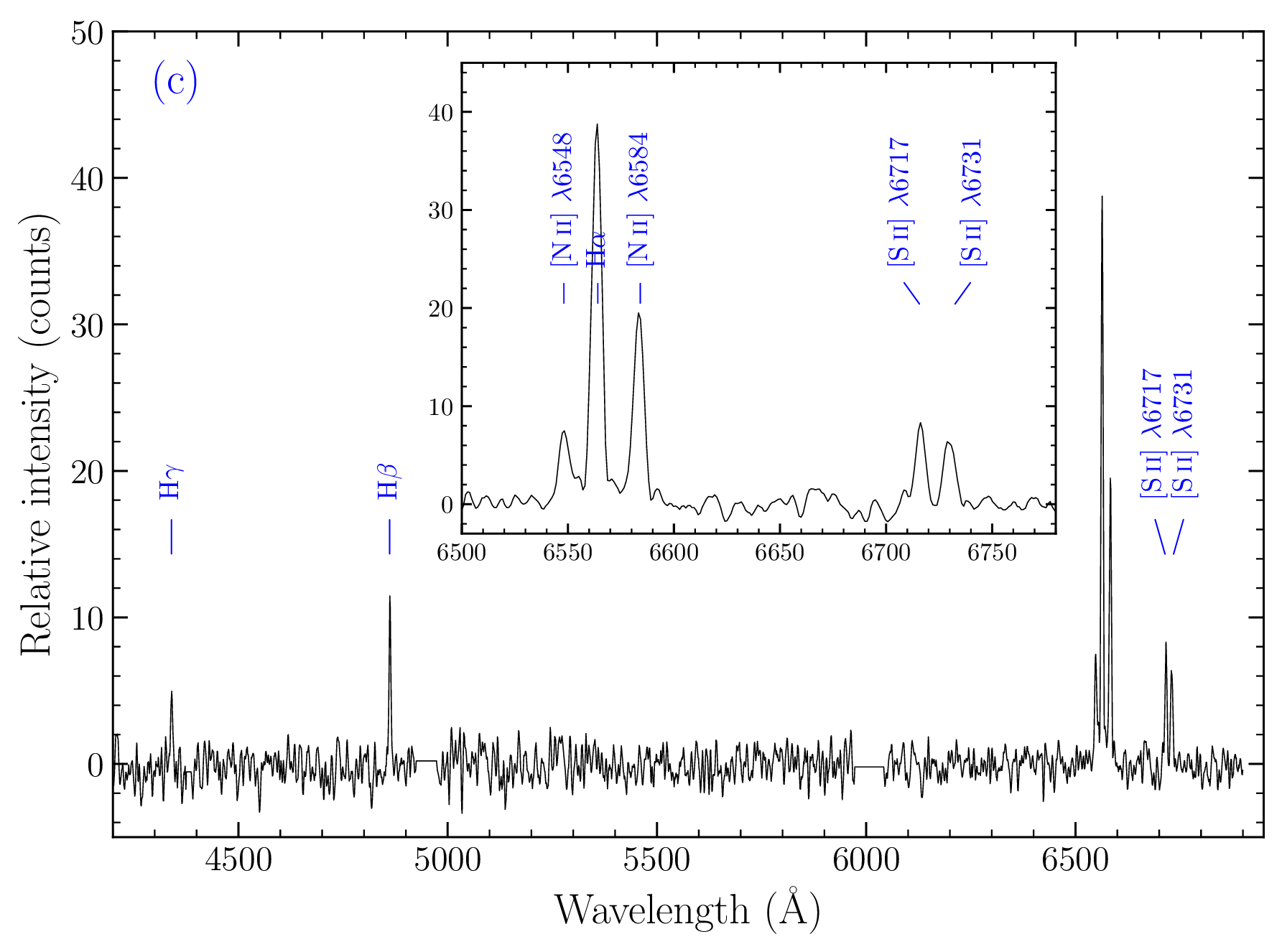}
	\includegraphics[width=8.5cm,clip=]{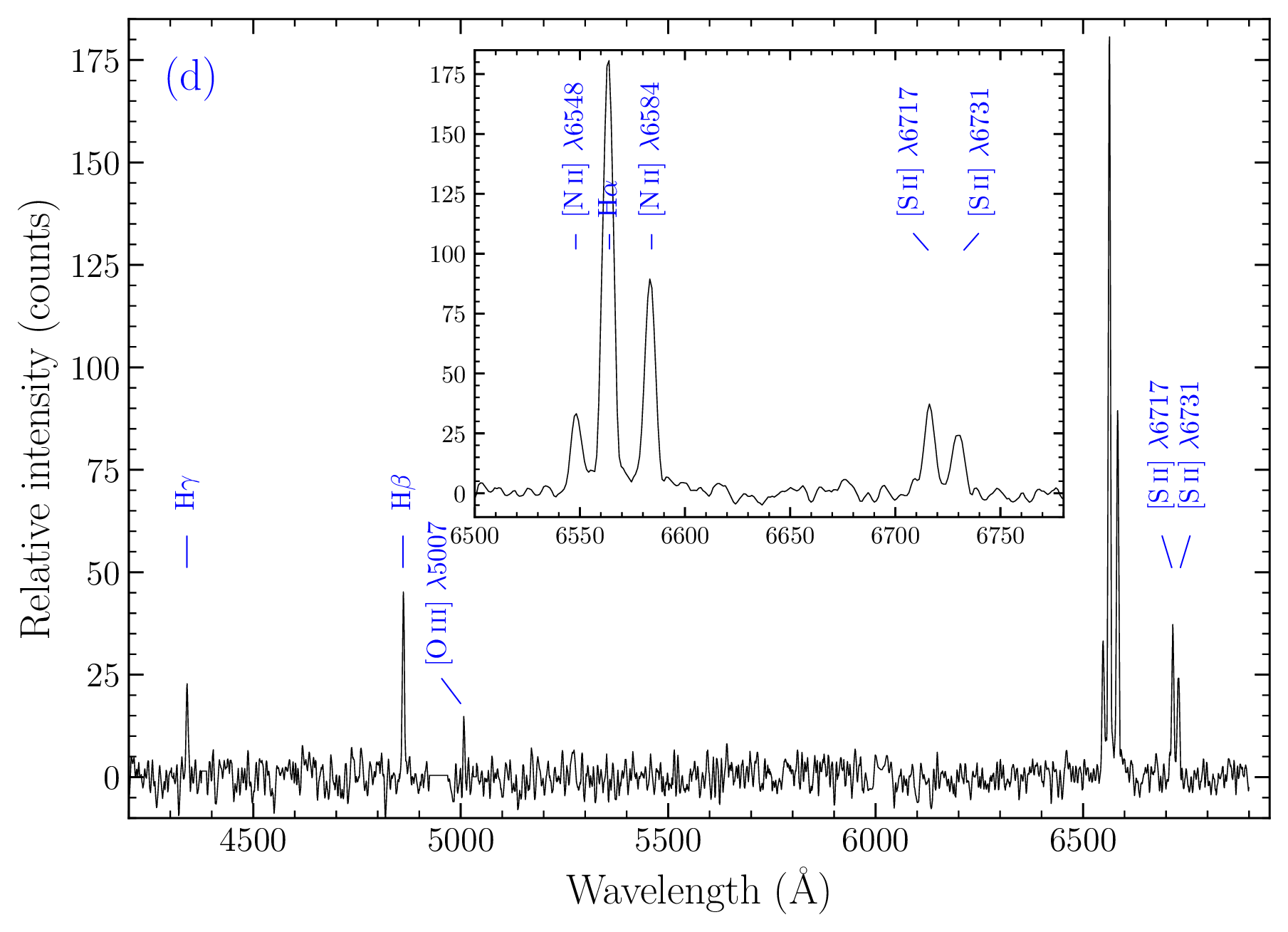}
	\caption{1D RSS spectra of the nebula (panel a), knot (panel b) and regions\,A and B (panels\,c 
	and d, respectively). The major emission lines are marked. The inserts show parts of the spectra 
	around the H\,$\alpha$ line. The gaps in the spectra at $\approx4900$ and 6000\,\AA \, are due to
	gaps between CCDs. See text for details. 
	}
	\label{fig:spectra}
	\end{center}
\end{figure*}

\section{RSS spectroscopy of the nebula}
\label{sec:neb-spec}

Figure\,\ref{fig:pro} shows the H\,$\alpha$ and [N\,{\sc ii}] $\lambda$6584 emission line 
intensity profiles along the RSS slit, oriented along the presumed axis of rotational symmetry of 
the nebula (PA=80$\degr$). The figure shows that HD\,93795 is surrounded by a region of strong 
nitrogen emission, extended between $\approx +20$ arcsec and $-15$ arcsec, i.e. within the angular 
extent of the IR nebula. We will refer to this region as the ``nebula". There is also another region 
of strong nitrogen emission to the west of the star between $\approx -30$ and $-40$ arcsec. 
Interestingly, this region corresponds to a knot of enhanced IR emission, most clearly visible in 
the IRAC 8\,$\micron$ image (see Fig.\,\ref{fig:neb}). We will refer to this region as the ``knot" 
and discuss it in Section\,\ref{sec:dis}. Besides, there are also two regions of bright 
H\,$\alpha$ emission to the west of the star, located between $\approx-18$ and $-30$ arcsec and 
$\approx-40$ and $-80$ arcsec; hereafter, respectively, ``region A" and ``region B". 

\begin{table*}
\centering{
\caption{Line intensities of the circumstellar nebula and three other regions along the RSS slit.
See text for details.}
\label{tab:neb}
\begin{tabular}{lcccccccc} 
\hline
 & \MC {2}{c}{Nebula} & \MC {2}{c}{Knot} & \MC {2}{c}{Region A} & \MC {2}{c}{Region B} \\
$\lambda_{0}$(\AA) Ion   & F($\lambda$)/F(H$\beta$) & I($\lambda$)/I(H$\beta$) & F($\lambda$)/F(H$\beta$) & 
I($\lambda$)/I(H$\beta$) & F($\lambda$)/F(H$\beta$)& I($\lambda$)/I(H$\beta$) & F($\lambda$)/F(H$\beta$) &
I($\lambda$)/I(H$\beta$)\\ 
\hline
4340\ H$\gamma$\        & 0.63$\pm$0.13 & 0.75$\pm$0.15 & 0.50$\pm$0.14 & 0.60$\pm$0.17 & 0.47$\pm$0.05 
& 0.52$\pm$0.06 & 0.45$\pm$0.04 & 0.53$\pm$0.04 \\
4861\ H$\beta$\         & 1.00$\pm$0.12 & 1.00$\pm$0.12 & 1.00$\pm$0.13 & 1.00$\pm$0.13 & 1.00$\pm$0.07 
& 1.00$\pm$0.07 & 1.00$\pm$0.06 & 1.00$\pm$0.06 \\
5007\ [O\ {\sc iii}]\   & -- & -- & -- & -- & -- & -- & 0.22$\pm$0.04 & 0.21$\pm$0.04 \\ 
6548\ [N\ {\sc ii}]\    & 1.35$\pm$0.14 & 0.90$\pm$0.10 & 1.02$\pm$0.13 & 0.64$\pm$0.09 & 0.71$\pm$0.06 
& 0.55$\pm$0.05 & 0.77$\pm$0.04 & 0.52$\pm$0.03 \\
6563\ H$\alpha$\        & 4.38$\pm$0.38 & 2.89$\pm$0.28 & 4.61$\pm$0.45 & 2.89$\pm$0.31 & 3.76$\pm$0.21 
& 2.89$\pm$0.18 & 4.30$\pm$0.19 & 2.89$\pm$0.14 \\
6584\ [N\ {\sc ii}]\    & 4.17$\pm$0.35 & 2.74$\pm$0.26 & 3.18$\pm$0.31 & 1.99$\pm$0.21 & 2.16$\pm$0.12 
& 1.66$\pm$0.10 & 2.35$\pm$0.10 & 1.57$\pm$0.08 \\
6717\ [S\ {\sc ii}]\    & 1.56$\pm$0.17 & 1.00$\pm$0.12 & 2.11$\pm$0.23 & 1.28$\pm$0.15 & 0.79$\pm$0.07 
& 0.60$\pm$0.05 & 0.89$\pm$0.05 & 0.58$\pm$0.04 \\
6731\ [S\ {\sc ii}]\    & 1.39$\pm$0.15 & 0.88$\pm$0.11 & 1.64$\pm$0.19 & 0.99$\pm$0.12 & 0.67$\pm$0.06 
& 0.50$\pm$0.05 & 0.64$\pm$0.05 & 0.41$\pm$0.03 \\
  & & \\
$C$(H $\beta$)          & \MC {2}{c}{0.55$\pm$0.11} & \MC {2}{c}{0.61$\pm$0.12} & \MC {2}{c}{0.34$\pm$0.07}
& \MC {2}{c}{0.52$\pm$0.06}\\  
$E(B-V)$                & \MC {2}{c}{0.37$\pm$0.07} & \MC {2}{c}{0.41$\pm$0.08} & \MC {2}{c}{0.23$\pm$0.05} 
& \MC {2}{c}{0.35$\pm$0.04} \\
$n_{\rm e}$([Si\,{\sc ii}])     & \MC {2}{c}{$340^{+420} _{-240} \, {\rm cm}^{-3}$} & \MC {2}{c}{$130^{+325} 
_{-120} \, {\rm cm}^{-3}$} & \MC {2}{c}{$250^{+265} _{-175} \, {\rm cm}^{-3}$} & \MC {2}{c}{$25^{+135} _{-15} 
\, {\rm cm}^{-3}$} \\
\hline
\end{tabular}
}
\end{table*}

In Fig.\,\ref{fig:spectra} we present spectra of the above four regions. All of them show Balmer 
emission lines and two strong forbidden emission doublets of [N\,{\sc ii}] $\lambda\lambda$6548,
6583 and [S\,{\sc ii}] $\lambda\lambda$6716, 6731. The spectrum of the region\,B also clearly shows 
the presence of the [O\,{\sc iii}] $\lambda$5007 emission lines. In Table\,\ref{tab:neb} we give 
for each region the observed intensities of the detected lines normalized to H\,$\beta$, 
F($\lambda$)/F(H\,$\beta$), the reddening-corrected line intensity ratios, I($\lambda$)/I(H\,$\beta$), 
the logarithmic extinction coefficient, $C$(H\,$\beta$), the colour excess $E(B-V)$, and the electron 
number density $n_{\rm e}$([S\,{\sc ii}]) derived from the [S\,{\sc ii}] $\lambda\lambda$6716, 6731 
line ratio. The lines in Table\,\ref{tab:neb} were measured with program described in Kniazev et al. 
(2004). 

One can see that the [S\,{\sc ii}] doublet is particularly strong in the spectra of the nebula and the 
knot, which suggests that their emission is the result of shock excitation. Indeed, using the 
$\log$(H\,$\alpha$/[N\,{\sc ii}]) versus $\log$(H\,$\alpha$/[S\,{\sc ii}]) diagnostic diagram
(e.g. Sabbadin, Minello \& Bianchini 1977; Kniazev, Pustilnik \& Zucker 2008), one finds that the 
nebula and the knot are located in the same area as supernova remnants, whose optical emission is
due to shock heating. Similarly, the line ratios in regions\,A and B place them in the area occupied 
by \hii regions. 

Note that the colour excess towards the nebula is lower by $\approx0.5$ mag than that derived 
for HD\,93795 (see Section\,\ref{sec:lum}). We attribute this difference to internal extinction in 
the ridge, which we interpret as a disc-like structure viewed edge-on (see next section). On the other 
hand, the colour excesses measured for all four regions along the slit are consistent with each other 
within the margins of errors, and are within the range of colour excesses found for stars in the Carina 
Nebula (e.g. Herbst 1976; Forte 1978). This could be considered as an indication that HD\,93795 is 
located in the Carina Nebula as well.
      
\section{Discussion}
\label{sec:dis}

The results of our spectral analysis indicate that HD\,93795 is one more example of a massive star 
which managed to produce a compact circumstellar nebula soon after it left the main sequence. 
We interpret the bright ridge within the nebula as a flattened (disc-like) structure viewed edge-on. 
If correct, this interpretation would suggest that rotation have played a role in shaping the nebula. 
If HD\,93795 was born as a single star, then one needs to assume that it was, at least at the time of 
formation of the nebula, a very fast rotator. In this case, the axial symmetry axis of the nebula 
is defined by the rotational axis of the star. Moreover, the measured low projected rotational velocity
of HD\,93795 would be equal to the true rotational velocity, meaning that the star has significantly spun 
down since the formation of the nebula. Alternatively, HD\,93795 might be a product of binary evolution, 
and in this case the symmetry axis of the nebula would coincide with the angular momentum vector of the 
binary orbit. Let us discuss possible scenarios for the origin of axially-symmetric nebulae around 
relatively unevolved single and binary stars. 

In the early stages of a single massive star's evolution the formation of an axially-symmetric circumstellar 
nebula could be related to the bi-stability jump (Pauldrach \& Puls 1990; Lamers \& Pauldrach 1991), which is 
manifested in drastic changes in the wind properties caused by decrease of the stellar effective temperature 
below some critical value ($\approx21\,000-25\,000$\,K; Vink, de Koter \& Lamers 1999). Recent bi-stability 
models by Vink (2018) predict that decrease of $T_{\rm eff}$ from $\approx25\,000$ to 17\,000\,K is 
accompanied by about an order of magnitude increase in $\dot{M}$ and by about the same magnitude decrease 
in $v_\infty$, leading to a factor of $\sim100$ increase in the density of the stellar wind. 
In fast-rotating stars $T_{\rm eff}$ is lower near the stellar equator because of the effect of 
gravity-darkening. In these stars the bi-stability jump first takes place at low latitudes, which might 
result in a dense axially-symmetric equatorial outflow (Lamers \& Pauldrach 1991; Lamers, Snow \& Lindholm 
1995).

The main problem with this scenario is that the size of nebulae created by the rotationally induced 
bi-stability jump mechanism should be much larger than the size of the nebula around HD\,93795. 
Indeed, the stellar evolutionary models (e.g. Ekstr\"om et al. 2012) show that in fast-rotating stars 
with the initial mass in the range derived for HD\,93795 the effective temperature decreases 
from 25\,000 to 17\,000\,K on the time scale from several 1000 to $\approx10\,000$\,yr. During this time, 
the equatorial wind will spread over a distance from several to about 10 pc, which is an order of magnitude 
larger than the radius of the nebula around HD\,93795 of $\approx0.36(d/5 \, {\rm kpc})$\,pc. 
In this connection, we note that the radii of equatorial rings of all five previously known bipolar 
nebulae produced 
by blue supergiants (see table\,7 in Gvaramadze et al. 2015 for main properties of these stars and their 
circumstellar rings) span a narrow range from $\approx0.2$ to 0.3 pc, and that their expansion velocities 
of $\approx10-30 \, \kms$  are about one or two orders of magnitude lower than the wind velocity of 
blue supergiants on the cool side of the bi-stability jump. This strongly suggests that the bi-stability 
jump mechanism cannot be responsible for the origin of these nebulae. Although the expansion 
velocity of the nebula around HD\,93795 is unknown, its small size also suggests that the bi-stability 
jump is not responsible for its origin. We therefore conclude that compact axially-symmetric nebulae hardly 
can be produced by single massive stars during their early post-main-sequence evolution. 

Among the five already known compact axially-symmetric nebulae (rings) produced by blue 
supergiants, the most studied
one is the equatorial ring around Sk$-69\degr$202 (Wampler et al. 1990; Jakobsen et al. 1991) --- the 
progenitor star of SN\,1987A. The proposed models for its origin converge on the idea that Sk$-69\degr$202 
was a binary system that merged shortly ($\sim10\,000$\,yr) before the supernova explosion (e.g. 
Podsiadlowski 1992; Collins et al. 1999; Morris \& Podsiadlowski 2007) to produce a highly flattened 
circumstellar structure slowly expanding in the system's orbital plane. Presumably, this structure is 
composed either of material expelled by the merging binary during the common-envelope phase (Morris 1981) 
or of wind material lost by the inflated merger product during its contraction into a blue supergiant 
(Heger \& Langer 1998).

The photospheric abundances of other four blue supergiants (Sher\,25, HD\,168625, [SBW2007]\,1 
and MN18) with equatorial rings indicate that these stars did not evolve through the red 
supergiant phase (e.g. Smartt et al. 2002; Smith et al. 2007; Gvaramadze et al. 2015), meaning that the 
rings were produced while the stars were in the main sequence phase or at the beginning of the blue 
supergiant phase (cf. Lamers et al. 2001). The luminosities of these stars range from $\log(L/\lsun)\approx4.7$ 
to 5.8, which indicates that the stellar luminosity (or initial mass of the star) is not a crucial factor 
in formation of circumstellar nebulae. The effective temperatures of these stars also span quite a large 
range from $\approx11\,000$ to 22\,000\,K (spectral types from B6 to B1). What units 
them is the similarity of the parameters (size, expansion velocity, number density) of their circumstellar 
rings (see table\,7 in Gvaramadze et al. 2015). This points to the possibility that perhaps there is a 
universal mechanism for formation of rings around massive stars. 

Since most of massive stars are formed in 
binary and multiple systems (e.g. Sana et al. 2012; Chini et al. 2012) and because massive binaries can 
merge both on the main sequence and in later stages of their evolution (e.g. de Mink et al. 2014), 
we suggest that the origin of the nebulae around HD\,93795 and the four other blue supergiants evolving 
redwards for the first time is also due to merger 
of binary systems. This implies that the equatorial rings should be short-lived objects because after 
thermal relaxation of the merger products the fast wind from the newly formed blue supergiants quickly 
disperse them. Assuming that all mergers result in the origin of axially-symmetric nebulae and that 
the lifetime of these nebulae is $\sim10,000$\,yr, one can show (see Gvaramadze et al. 2015) that there are
should be $\sim10$ such nebulae in the Milky Way.

During the initial stage of interaction between the fast wind and the slow dense material of 
the equatorial ejecta, the stellar wind will flow freely in the polar directions to create bipolar lobes.
The ``horns" on each side of the ridge in the nebula around HD\,93795 may represent the beginning of 
these lobes, while the strong nitrogen emission detected close to the star may come from the 
equatorial ejecta shocked by the stellar wind. If so, then the knot of shocked gas to the west of 
HD\,93795 may be a cloudlet in the interstellar medium, which was accidentally caught in the path of the 
stellar wind flowing in the polar direction.  

Finally, we discuss the possible membership of HD\,93795 of the Car\,OB1 association.
While taken at face value the {\it Gaia} parallax indicates that HD\,93795 is a factor of two more 
distant than Car\,OB1, the distance to HD\,93795 could be overestimated because of the systematic offset 
in the {\it Gaia} DR2 parallaxes. The large distance to HD\,93795 is also suggested by the flux-weighted 
gravity-luminosity relationship for blue supergiants, but the uncertainty inherent to this relationship 
does not exclude the possibility that the star is at a smaller distance. On the other hand, the colour 
excess towards the region of shocked gas in the close vicinity of HD\,93795 (we associate this material 
with the slow equatorial ejecta shocked by the fast stellar wind) is typical of stars in the Carina Nebula, 
which points to the possibility that HD\,93795 is located in the Carina Nebula as well. Although the existing 
data did not allow us to unambiguously constrain the distance to HD\,93795, we conclude that this star 
could be a member of the Car\,OB1 association. Hopefully, the forthcoming third {\it Gaia} data release 
will resolve this issue.

\section{Summary}
\label{sec:sum}

We have discovered a compact IR nebula around the emission-line star HD\,93795 using archival data from the 
{\it Spitzer Space Telescope}. Follow-up optical spectroscopy of HD\,93795 with the Southern African 
Large Telescope led to its classification as a B9\,Ia star, while modelling of its spectrum showed 
that HD\,93795 is a relatively unevolved star moving redward for the first time. These findings support 
the previous view that massive stars can produce circumstellar nebulae while they are still on the main
sequence or shortly after they left it. 

The IR nebula has a square shape with a bright ridge in the middle and two symmetric flanks perpendicular 
to the ridge. We interpret the ridge as an axially-symmetric flattened structure viewed edge-on, and suggest 
that the flanks outline the beginning of bipolar lobes. The axially-symmetric shape of the nebula suggests 
that HD\,93795 was either a single fast-rotating star or a close binary system, which has merged in the recent 
past. In the first case, the only possible way to produce an axially-symmetric nebula is through the 
rotationally induced bi-stability jump mechanism. The size and expansion velocity of such a nebula, however, 
would be much larger than those of the known circumstellar rings around blue supergiants. We therefore conclude, 
that the merger of binary systems appears to be a more natural process leading to the formation of compact 
slowly expanding and flattened circumstellar nebulae. 

\section{Acknowledgements}
This work is based on observations obtained with the Southern African Large Telescope (SALT), 
programmes 2013-1-RSA$\_$OTH-014, 2015-1-SCI-017, 2018-1-MLT-008, and collected at the European 
Southern Observatory under ESO programme 092.A-9020(A). V.V.G. acknowledges support from the 
Russian Science Foundation under grant 19-12-00383. A.Y.K. acknowledges support from the National 
Research Foundation (NRF) of South Africa. This research has made use of the NASA/IPAC Infrared 
Science Archive, which is operated by the Jet Propulsion Laboratory, California Institute of 
Technology, under contract with the National Aeronautics and Space Administration, the SIMBAD 
data base and the VizieR catalogue access tool, both operated at CDS, Strasbourg, France, and 
data from the European Space Agency (ESA) mission {\it Gaia} (https://www.cosmos.esa.int/gaia), 
processed by the {\it Gaia} Data Processing and Analysis Consortium 
(DPAC, https://www.cosmos.esa.int/web/gaia/dpac/consortium). Funding for the DPAC has been 
provided by national institutions, in particular the institutions participating in the Gaia 
Multilateral Agreement.

\end{document}